\documentclass[twocolumn,notitlepage]{revtex4-2}
\usepackage{amsmath}
\usepackage{amssymb}
\usepackage{graphicx}
\usepackage{subcaption}
\usepackage{dcolumn}
\usepackage{bm}
\usepackage{xcolor}
\usepackage{hyperref}
\usepackage{soul}

\newcommand{\iu}{{i\mkern1mu}}

\allowdisplaybreaks

\newcommand{\qed}{\nobreak \ifvmode \relax \else
	\ifdim\lastskip<1.5em \hskip- \lastskip
	\hskip 0.5em plus0em minus0.5em \fi \nobreak
	\vrule height0.75em width0.5em depth0.25em\fi}

\begin{document}

\title{A complete correspondence between the Newman-Penrose and 1+1+2 formalisms}

\author{Abbas M \surname{Sherif}$^{1}$}
\email{abbasmsherif25@gmail.com (corresponding author)}
 \affiliation{$^{1}$Institute of Mathematics, Henan Academy of Sciences (HNAS), 228 Mingli Road, Zhengzhou 450046, Henan, China.}

\author{Peter K S \surname{Dunsby}$^{2,3,4}$}
\email{peter.dunsby@uct.ac.za}
\affiliation{$^{2}$Department of Mathematics and Applied Mathematics, University of Cape Town, Rondebosch 7701, Cape Town, South Africa}
\affiliation{$^{3}$Center for Space Research, North-West University, Potchefstroom 2520, South Africa}
\affiliation{$^{4}$South African Astronomical Observatory, Observatory 7925, Cape Town, South Africa}

\date{\today}

\begin{abstract}
We establish a correspondence between the Newman-Penrose and 1+1+2 semitetrad covariant formalisms by expressing all Newman-Penrose spin coefficients, Ricci scalars, and Weyl scalars in terms of the scalar, vector, and tensor variables of the 1+1+2 decomposition. In addition, we provide some discussions on the correspondence between the gauge structures of the formalisms. This provides a direct dictionary between two widely used approaches to general relativity and gives a geometrical interpretation of Newman-Penrose quantities in terms of covariantly defined 1+1+2 variables. As a simple demonstration, we use this mapping to derive inequalities on the Newman-Penrose scalars and the cosmological constant, which constrains the existence of future outer trapping horizons in spacetimes exhibiting local rotational symmetry. 
\end{abstract}

\keywords{Black hole horizons; Conformal Symmetry; Marginally outer trapped surfaces (MOTS); MOTS stability}

\maketitle


\section{\label{sec:level1}Introduction}


The broad utility of the Newman-Penrose (NP) formalism \cite{pen1,pen2} has been demonstrated through its extensive application in gravitational perturbation theory (see, for example, \cite{chand1}) and in the study of black hole horizons and their perturbations \cite{ash1,ash2,ash3}. The introduction of the 1+1+2 semi-tetrad covariant formalism, on the other hand, has provided a powerful framework for investigating gravitational systems with preferred spatial directions and inhomogeneities \cite{chris1,chris2}. 

The two formalisms are gauge-equivalent descriptions of the same underlying spacetime, although they emphasize different structures. In particular, the difference is primarily in representation. The Newman-Penrose formalism emphasizes the full Lorentz gauge symmetry and is particularly powerful in algebraically special spacetimes and radiation problems. The 1+1+2 semitetrad formalism emphasizes geometric adaptation by fixing much of the Lorentz freedom at the outset, leaving only the intrinsic $SO(2)$ sheet rotation as a residual frame gauge. The two approaches are therefore complementary: the NP formalism provides maximal gauge flexibility, while the 1+1+2 formalism trades that flexibility for variables that are more directly adapted to the physical symmetries of the spacetime and often more transparent in applications involving preferred timelike and spatial directions. This complementarity means that for a given spacetime, depending on the particular problem one seeks to address, we can switch between the approaches.

The strength of the NP formalism lies in its elegance, computational efficiency, and broad range of applications. Beyond gravitational perturbation theory, it has played a central role in the theory of (weakly) isolated horizons, providing a geometric characterization of black holes in equilibrium \cite{ash100,ash101,ash102,ash1,ash2,ash3}. These horizons are null hypersurfaces foliated by co-dimension one surfaces with vanishing outward null expansion, and their mechanics, geometry, and existence have been studied extensively within the NP framework. The formalism has subsequently been employed in a variety of related investigations, including studies of the Petrov type D equation on isolated null surfaces \cite{dob1,kam1} and its extension to higher-genus horizons \cite{dob2}, analyses of the near-horizon geometry equation \cite{dob3}, and investigations of the spacetime geometry in the neighbourhood of non-extremal weakly isolated horizons \cite{krish1}.

The 1+1+2 formalism, however, possesses several distinctive advantages. First, the Einstein field equations can be expressed as a first-order system governing curvature and dynamical variables along preferred spacetime directions, supplemented by a set of constraints, rather than the second-order systems that arise in coordinate-based approaches. Second, the variables arising from the decomposition possess direct geometric and physical interpretations, being naturally associated with the temporal and spatial congruences. Third, in perturbation theory, gauge invariance follows directly from the Stewart-Walker lemma \cite{stew1}. The second feature is particularly significant, as it allows geometric meaning to be attached to structures arising in alternative formulations of general relativity. This has motivated, for example, the application of the 1+1+2 formalism to the study of horizons in exact spacetimes \cite{ellis1,abb1,abb2}. In \cite{abb3}, the response of a null horizon to linear perturbations was also investigated within this framework. It is therefore natural to expect that establishing a direct correspondence between the NP and 1+1+2 formalisms may provide new insights into gravitational systems, particularly by clarifying the geometric interpretation of horizon-related quantities formulated in the language of the NP approach.

In the present work, however, we derive the complete set of relations for all NP spin coefficients, Ricci scalars, and Weyl scalars. To the best of our knowledge, this constitutes the first complete correspondence between the NP and 1+1+2 formalisms. In doing so, we construct a direct dictionary between two widely used approaches to general relativity. As a simple illustration of the utility of this correspondence, we derive necessary conditions for the existence of future outer trapping horizons, expressed as inequalities involving NP curvature scalars and the cosmological constant.

Some aspects of this correspondence have previously been obtained in more restricted settings. For example, Pratten \cite{prat1} exploited relations between the two formalisms in the study of gravitational perturbations, where perturbation variables were related to the electric and magnetic parts of the Weyl tensor and connected to the NP Weyl scalars $\Psi_0$, $\Psi_2$, and $\Psi_4$. This provided an elegant framework for analysing perturbations of Schwarzschild spacetime. Similarly, in \cite{rit1}, the Ricci NP scalars were computed for a restricted subclass of locally rotationally symmetric spacetimes in the context of energy transfer between matter and the free gravitational field.

As an illustration of the usefulness of the resulting dictionary, we apply it to the study of black hole horizons in locally rotationally symmetric spacetimes. In particular, we obtain necessary conditions for the existence of future outer trapping horizons expressed entirely in terms of NP curvature scalars and the cosmological constant.

The paper is organized as follows. In Section \ref{2}, we briefly review the 1+1+2 semitetrad covariant formalism and establish the notation and conventions used throughout. In Section \ref{3}, we derive the complete correspondence between the NP and 1+1+2 variables. In Section \ref{4}, we apply these relations to obtain existence criteria for black hole horizons in locally rotationally symmetric spacetimes. Finally, in Section \ref{5}, we summarize our results and discuss possible future applications of the correspondence established here.

Throughout this work, $\left(-,+,+,+\right)$ will be the metric signature we use. The spacetime volume form is denoted by $\eta_{abcd}$, while $\bar{\eta}_{abc}=\eta_{dabc}u^d$ is the spatial volume form orthogonal to $u^a$, and $\tilde{\eta}_{ab}=\bar{\eta}_{abc}e^c$ is the area form on the two-sheet. Overbars on projected 1+1+2 quantities denote sheet projection, while an overbar on a complex quantity denotes complex conjugation. Tildes on Newman--Penrose quantities are used to distinguish spin coefficients from similarly named 1+1+2 variables.


\section{A brief review of the 1+1+2 formulation}\label{2}

For completeness, we briefly review the 1+1+2 semitetrad covariant formulation of general relativity and establish the notation
used throughout this work. We curtail many of the intricate details as there is a wealth of literature which comprehensively introduces the formulation \cite{chris1,chris2}, which we shall follow. To begin with, one considers a spacetime with metric tensor $g_{ab}$ and compatible covariant derivative $\nabla_a$, foliated into spacelike slices, and suppose these slices are themselves foliated by ``surfaces'' (these spaces/pseudo-surfaces are referred to as sheet in the literature since in general they are rather a collection of tangent planes, and only under certain conditions they are what is referred to as  a genuine surface). The surfaces have unit normal $e^a$, orthogonal to the timelike direction, which allows us to decompose the spacetime metric as
\begin{align}
q_{ab}=h_{ab}-e_ae_b=g_{ab}+u_au_b-e_ae_b,
\end{align}
where $q_{ab}$ and $h_{ab}$ are respectively the projection tensors onto the sheet and the 3-space orthogonal to $u^a$.

A 3-vector $\psi_a$ splits as
\begin{align}
\psi_a=\bar\psi e^a+\bar\psi^a,\quad\bar\psi=\psi_ae^a,\quad\bar\psi^a=q^b_{\ a}\psi_b.\label{3v}
\end{align}

Similarly, a projected, symmetric, trace-free 3-tensor \(\Psi_{ab}\)
may be decomposed as
\begin{align}\label{split3}
\Psi_{ab}
=
\bar{\Psi}
\left(e_a e_b-\frac{1}{2}q_{ab}\right)
+2\bar{\Psi}_{(a}e_{b)}
+\bar{\Psi}_{ab},
\end{align}
where
\begin{align*}
\bar{\Psi}&=\Psi_{ab}e^ae^b,\quad\bar{\Psi}_a=q_a{}^b\Psi_{bc}e^c,\\
\bar{\Psi}_{ab}&=\left(q_{(a}{}^c q_{b)}{}^d-\frac12 q_{ab}q^{cd}\right)\Psi_{cd}.
\end{align*}

Three directional derivatives result from the decomposition:
\begin{itemize}
\item[(i)] Along $u^a$ (``dot'' derivative): \\$\dot{\psi}^{a\cdots b}_{\ \ \ c\cdots d}=u^f\nabla_f\psi^{a\cdots b}_{\ \ \ c\cdots d}$,
\item[(ii)] Along $e^a$ (``prime'' derivative): \\$(\psi^{a\cdots b}_{\ \ \ c\cdots d})'=e^f\nabla_f\psi^{a\cdots b}_{\ \ \ c\cdots d}$,
\item[(iii)] Along the sheet (``sheet'' derivative):\\ $\mathcal{D}_f\psi^{a\cdots b}_{\ \ \ c\cdots d}=q^{\bar a}_{\ a}\cdots q^{\bar b}_{\ b}q^{\bar c}_{\ c}\cdots q^{\bar d}_{\ d}q^e_{\ f}\nabla_e\psi^{\bar a\cdots \bar b}_{\ \ \ \bar c\cdots \bar d}$,
\end{itemize}
for any tensor $\psi^{a\cdots b}_{\ \ \ c\cdots d}$, with $\mathcal{D}_a$ denoting the derivative compatible with the 2-metric $q_{ab}$. Finally, a scalar $\psi$ will have its gradient decomposed as
\begin{align}
\nabla_a\psi=-\dot{\psi}u_a+\psi'e_a+\mathcal{D}_a\psi.
\end{align}
Moving on to the geometry, the energy momentum tensor, under the splitting, decomposes as
\begin{align}
T_{ab}=\varrho u_au_b+ph_{ab}+2q_{(a}u_{b)}+\pi_{ab},
\end{align}
with the 3-vector $q_a$ and 3-tensor $\pi_{ab}$ decomposing according to \eqref{3v} and \eqref{split3}, so that from the Einstein Field equations 
\begin{align}
R_{ab}-\frac{1}{2}(R-2\Lambda)g_{ab}=T_{ab},
\end{align}
one obtains the Ricci tensor as
\begin{align}
R_{ab}&=\frac{1}{2}(\varrho+3p-2\Lambda)u_au_b+\frac{1}{2}(\varrho-p+2\Lambda+2\Pi)e_ae_b\nonumber\\
&+\frac{1}{2}(\varrho-p+2\Lambda-\Pi)q_{ab}+2(Qu_{(a}+\Pi_{(a})e_{b)}\nonumber\\
&+2Q_{(a}u_{b)}+\Pi_{ab}.
\end{align}
The various scalars, vectors and tensor introduced above have the following definitions: $\varrho=T_{ab}u^au^b$ is the (local) energy density, $3p=-T_{ab}h^{ab}$ is the (isotropic) pressure, $q_a=h^b_{\ a}T_{bc}u^c$ is the heat 3-flux vector, and $\pi_{ab}$ captures deviation from isotropy.

The Weyl tensor takes the decomposed form
\begin{align}\label{wt1}
C_{abcd}&=4u_{[c}u^{[a}E^{b]}_{\ d]}+4E_{[c}^{\ [a}h^{b]}_{\ d]}\nonumber\\
&-2\bar\eta^{abf}H_{f[c}u_{d]}-2\bar\eta_{cd}^{\ \ f}H_{f}^{\ [a}u^{b]},
\end{align}
where the square brackets indicate antisymmetrization, and the 3-tensors
\begin{align}\label{wt1}
E_{ab}=C_{acbd}u^cu^d,\quad H_{ab}=\frac{1}{2}\bar\eta_a^{\ ef}C_{efbd}u^d,
\end{align}
respectively represent the \textit{electric} and \textit{magnetic} parts of the Weyl tensor. These are 3-tensors that can be appropriately decomposed according to \eqref{split3}.

The evolution and propagation equations of $u^a$ and $e^a$ are
\begin{align}
\dot{u}^a&=\mathcal{A}e^a+\mathcal{A}^a,\nonumber\\
\hat{u}^a&=\left(\frac{1}{3}\theta+\Sigma\right)e^a+\left(\Sigma^a+\tilde\eta^{ab}\Omega_b\right),\\
\dot{e}^a&=\mathcal{A}u^a+\alpha^a,\quad\hat{e}^a=a^a,
\end{align}
and their full covariant derivatives are given by
\begin{align}
\nabla_au_b&=-u_a\left(\mathcal{A}e_b+\mathcal{A}_b\right)+\left(\frac{1}{3}\theta+\Sigma\right)e_ae_b\nonumber\\
&+\frac{1}{2}\left(\frac{2}{3}\theta-\Sigma\right)q_{ab}+\Omega\tilde\eta_{ab}+\Sigma_{ab}\label{cd1}\\
&+2\left(e_{(a}\Sigma_{b)}+e_{[a}\tilde\eta_{b]c}\Omega^c\right),\nonumber\\
\nabla_ae_b&=-u_a\left(\mathcal{A}u_b+\alpha_b\right)+\left(\frac{1}{3}\theta+\Sigma\right)e_au_b\nonumber\\
&+\frac{1}{2}\phi q_{ab}+\xi\tilde\eta_{ab}+\zeta_{ab}+e_aa_b\label{cd2}\\
&+\left(\Sigma_a-\tilde\eta_{ac}\Omega^c\right)u_b.\nonumber
\end{align}
$\mathcal{A}$ is the acceleration scalar, $\mathcal{A}_a$ the acceleration 2-vector, $\theta=h^{ab}\nabla_au_b$ is the expansion, $\phi=q^{ab}\nabla_ae_b$ is referred to as the sheet/surface expansion, $2\Omega=\tilde\eta^{ab}\nabla_au_b$, $2\xi=\tilde\eta^{ab}\nabla_ae_b$ are the respective vorticities of $u^a$ and $e^a$, (there is a 2-vector, the vorticity 2-vector $\Omega_a$, which results from decomposing the usual vorticity 3-vector associated to $u^a$ according to the first relation of \eqref{3v}), $\Sigma,\Sigma_a,\Sigma_{ab}$ are scalar, 2-vector, and 2-tensor obtained by decomposing the 3-shear $\sigma_{ab}$ of $u^a$ according to \eqref{split3}, and $\zeta_{ab}=\mathcal{D}_{\{a}e_{b\}}$ is the shear of $e^a$, with curly brackets indicating fully projected and trace-free with respect to $q_{ab}$. 

The Einstein field equations can then be written as a set of first order equations in the convective derivatives and constraints for the covariant quantities appearing here, using the Ricci identities for $u^a$ and $e^a$
\begin{align}
2\nabla_{[a}\nabla_{b]}u_c=R_{abc}^{\ \ \ d}u_d,\quad 2\nabla_{[a}\nabla_{b]}e_c=R_{abc}^{\ \ \ d}e_d.
\end{align} 
We will not write down this huge set of equations as they are not needed for the current purpose of this work, but there is no shortage of papers in the literature containing and systematically deriving them \cite{chris1,chris2}. 

In the NP formalism, one typically constructs gauge -- invariant quantities by exploiting special tetrad choices (for example, principal null tetrads) or by identifying combinations of Weyl scalars that are invariant under the residual tetrad gauge. The Weyl variables $\Psi_0$ and $\Psi_4$ in vacuum Petrov type D backgrounds are classic examples.

In the 1+1+2 formalism, the Stewart-Walker lemma provides a natural route to gauge invariance: variables that vanish on the background are automatically gauge invariant at first order. Because highly symmetric backgrounds such as Schwarzschild or LRS spacetimes admit many vanishing sheet vectors and tensors, the formalism naturally yields a large set of gauge-invariant perturbation variables. In such cases, the 1+1+2 formulation proves more effective.

A simple example where this preference of approach is transparent is the perturbations of the Schwarzschild spacetime. The spacetime possesses a canonical geometric splitting: a preferred timelike direction given by the static Killing observers and a preferred radial direction, specifying the 2-spheres of symmetry. This is precisely the 1+1+2 decomposition. The $u^a$ specifies our static observers and $e^a$ provides the radial direction. The sheet is then the tangent space of the symmetrical 2-spheres. Thus the geometry itself almost completely fixes the semitetrad gauge. The background is substantially simple: all sheet vectors and tensors vanish, and the only remaining non-zero background quantities are
\begin{align}
\{\mathcal{A},\phi,\mathcal{E}\},
\end{align}
providing automatic gauge invariance. Now suppose we perturb the Schwarzschild spacetime. Generically, all quantities that vanish in the background, including some scalars, sheet vectors, and fully projected and symmetric trace-free tensors, will give rise to perturbation variables. From the Stewart-Walker lemma it follows that the variable $\delta X$ for any quantity $X$ vanishing in the background is gauge-invariant. 

The work \cite{chris1} showed that the entire perturbation problem can be reduced to covariant master variables. For example, one introduces a transverse-traceless sheet tensor $W_{ab}$, constructed from the electric Weyl tensor, whose harmonic decomposition reproduces the well-known Regge-Wheeler equation. Thus, the Regge-Wheeler variable appears as the harmonic coefficient of a geometrically defined tensor with no coordinate gauge fixing is involved.

In the NP formalism on the other hand, one first fixes a null tetrad, usually the so-called Kinnersley tetrad, with a non-trivial gauge fixing. The Weyl scalars  are
\begin{align}
\Psi_2\neq0,\quad \Psi_j=0\quad j\neq2.
\end{align}

Perturbations are then given by the variables
\begin{align}
\delta\Psi_0,\quad \delta\Psi_4,
\end{align}
which obey the Teukolsky equation.While this approach is elegant and powerful, there are complications:
\begin{enumerate}
\item the perturbation variables are complex;
\item they depend on the null tetrad;
\item residual spin-boost freedom must be controlled;
\item reconstructing the metric from curvature perturbations is nontrivial (the metric reconstruction problem).
\end{enumerate}
By contrast, in the 1+1+2 formalism one works directly with covariant geometric quantities adapted to the spherical symmetry, and the relation to metric perturbations is often more transparent.

It follows that the perturbation equations in the 1+1+2 formalism resemble transport equations along physically distinguished directions, which is particularly advantageous for studies of gravitational collapse, horizon dynamics, and cosmological perturbations.


\section{Mapping the variables}\label{3}


In this section we derive the complete correspondence
between the NP and 1+1+2 formalisms by expressing all
NP spin coefficients, Ricci scalars and Weyl scalars
in terms of the scalar, vector and tensor variables of
the 1+1+2 decomposition.

We begin by constructing the real components of the null pair $\{\ell^a,n^a\}$, defining tangent vectors to null geodesics, from the unit directions $u^a$ and $e^a$ as
\begin{align}
\ell^a=\frac{1}{\sqrt{2}}(u^a+e^a),\quad n^a=\frac{1}{\sqrt{2}}(u^a-e^a),
\end{align}
cross normalized to $\ell_an^a=-1$. In the N-P formalism, a pair of complex conjugate vectors $\{m^a,\bar m^a\}$ satisfying
\begin{align}
m_am^a=\bar m_a\bar m^a=0,\quad m_a\bar m^a=1
\end{align}
completes the full tetrad. Then, the 2-sheet metric $q_{ab}$ and the complex pair $m^a$ and $\bar m^a$ are related as
\begin{align}
q_{ab}=2m_{(a}\bar m_{b)},
\end{align}
so that the spacetime metric decomposes as
\begin{align}
g_{ab}=-2\ell_{(a}n_{b)}+2m_{(a}\bar m_{b)},
\end{align}
The area 2-form $\tilde\eta_{ab}$ associated to $q_{ab}$ will then take the form 
\begin{align}
\tilde\eta_{ab}=2\iu \bar m_{[a}m_{b]},
\end{align}
with $\iu^2=-1$.

The NP scalars are defined by appropriate projections of the covariant derivatives of the null vectors, and contractions of the Ricci and Weyl tensors. The spin coefficients are given by

\begin{align}
\tilde\kappa&=-m^a\ell^b\nabla_b\ell_a,\quad\tilde\sigma=-m^am^b\nabla_b\ell_a,\\
\tilde\nu&=\bar m^an^b\nabla_bn_a,\quad\tilde\lambda=-\bar m^a\bar m^b\nabla_bn_a,\\
\tilde\tau&=-m^an^b\nabla_b\ell_a,\quad\tilde\rho=-m^a\bar m^b\nabla_b\ell_a,\\
\tilde\pi&=\bar m^a\ell^b\nabla_bn_a,\quad\tilde\mu=-\bar m^am^b\nabla_bn_a,\\
\tilde\alpha&=-\frac{1}{2}(n^a\bar m^b\nabla_b\ell_a-\bar m^a\bar m^b\nabla_bm_a),\\
\tilde\beta&=-\frac{1}{2}(n^am^b\nabla_b\ell_a-\bar m^am^b\nabla_bm_a),\\
\tilde\gamma&=-\frac{1}{2}(n^an^b\nabla_b\ell_a-\bar m^an^b\nabla_bm_a),\\
\tilde\epsilon&=-\frac{1}{2}(n^a\ell^b\nabla_b\ell_a-\bar m^a\ell^b\nabla_bm_a).
\end{align}
The Ricci NP scalars are given by
\begin{align}
\Phi_{00}&=\frac{1}{2}R_{ab}\ell^a\ell^b,\quad\Phi_{11}=\frac{1}{4}R_{ab}(\ell^an^b+m^a\bar m^b),\\
\Phi_{22}&=\frac{1}{2}R_{ab}n^an^b,\quad\Phi_{01}=\bar\Phi_{10}=-\frac{1}{2}R_{ab}\ell^am^b,\\
\Phi_{02}&=\bar\Phi_{20}=\frac{1}{2}R_{ab}m^am^b,\\
\Phi_{12}&=\bar\Phi_{21}=\frac{1}{2}R_{ab}\bar m^an^b
\end{align}
Finally, the Weyl N--P scalars are given by
\begin{align}
\Psi_0&=C_{abcd}\ell^am^b\ell^cm^d,\quad\Psi_1=C_{abcd}\ell^an^b\ell^cm^d,\\
\Psi_2&=C_{abcd}\ell^am^b\bar m^cn^d,\quad\Psi_3=C_{abcd}\ell^an^b\bar m^cn^d,\\
\Psi_4&=C_{abcd}n^a\bar m^bn^c\bar m^d.
\end{align}

So far these are the needed ingredients for the computations. We now compute all of the NP quantities in terms of the 1+1+2 quantities.

\subsection{Spin coefficients}

The NP spin coefficients are calculated as
\begin{align}
\tilde\kappa&=-\frac{1}{2}m_aY^a,\label{npspin1}\\
\tilde\sigma&=-\frac{1}{\sqrt{2}}(\Sigma_{ab}+\zeta_{ab})m^am^b,\label{npspin2}\\
\tilde\nu&=-\frac{1}{2}(2(\mathcal{A}_a+a_a)-Y_a)\bar m^a,\label{npspin3}\\
\tilde\lambda&=-\frac{1}{\sqrt{2}}(\Sigma_{ab}-\zeta_{ab})\bar m^a\bar m^b,\label{npspin4}\\
\tilde\tau&=-\tilde\kappa+(\Sigma_a+\tilde\eta_{ab}\Omega^b)m^a,\label{npspin5}\\
\tilde\rho&=-\frac{1}{2}\theta_{(\ell)}\nonumber\\
&-\frac{1}{\sqrt{2}}((\Omega+\xi)\tilde\eta_{ab}-(\Sigma_{ab}-\zeta_{ab}))\bar m^a m^b,\label{npspin6}\\
\tilde\pi&=\tilde\nu+(\mathcal{A}_a-\alpha_a)\bar m^a,\label{npspin7}\\
\tilde\mu&=-\frac{1}{2}\theta_{(n)}\nonumber\\
&-\frac{1}{\sqrt{2}}((\Omega-\xi)\tilde\eta_{ab}-(\Sigma_{ab}+\zeta_{ab}))m^a\bar m^b,\label{npspin8},\\
\tilde\alpha&=-\frac{1}{2}\left(\mathring{m}_a-\frac{1}{2}(\Sigma_a-\tilde\eta_{ab}\Omega^b)\right)\bar m^a,\\
\tilde\beta&=-\frac{1}{2}\left(\check{m}_a-\frac{1}{2}(\Sigma_a-\tilde\eta_{ab}\Omega^b)\right)m^a,\\
\tilde\gamma&=\frac{1}{2\sqrt{2}}\left(\mathcal{A}-\left(\frac{1}{3}\theta+\Sigma\right)+(\dot{m}_a-\hat{m}_a)\bar m^a\right),\\
\tilde\epsilon&=\frac{1}{2\sqrt{2}}\left(\mathcal{A}+\left(\frac{1}{3}\theta+\Sigma\right)+(\dot{m}_a+\hat{m}_a)\bar m^a\right),
\end{align}
where we have defined
\begin{align}
Y_a&=\mathcal{A}_a+\alpha_a+\Sigma_a+\tilde\eta_{ab}\Omega^b+a_a,\\
\theta_{(\ell)}&=q^{ab}\nabla_a\ell_b=\frac{1}{\sqrt{2}}(\frac{2}{3}\theta-\Sigma+\phi)=-2\Re(\tilde\varrho),\label{expan1}\\
\theta_{(n)}&=q^{ab}\nabla_an_b=\frac{1}{\sqrt{2}}(\frac{2}{3}\theta-\Sigma-\phi)=-2\Re(\tilde\mu)\label{expan2},
\end{align}
with the $\mathring{*}$ and $\check{*}$ derivatives being the directional derivatives along $\bar m^a$ and $m^a$, respectively, and where $\Re(\ )$ denotes ``the real part of''.

The NP spin coefficients are the Ricci rotation coefficients expressed in a null basis, whereas the kinematical variables of the 1+1+2 formalism, including $\mathcal{A},\theta,\Sigma,\phi,\xi$, and $\zeta_{ab}$ are precisely the same connection coefficients expressed in the semitetrad basis adapted to $\{u^a,e^a\}$.

\subsection{NP Ricci scalars}

The NP Ricci scalars are calculated as
\begin{align}
\Phi_{00}&=\frac{1}{4}(\varrho+p+\Pi-2Q),\label{npricci1}\\
\Phi_{11}&=\frac{1}{4}(\varrho-\frac{1}{2}\Pi-\Lambda+\Pi_{ab}m^a\bar m^b),\label{npricci2}\\
\Phi_{22}&=\frac{1}{4}(\varrho+p+\Pi+2Q),\label{npricci3}\\
\Phi_{01}&=-\frac{1}{2\sqrt{2}}(\Pi_a-Q_a)m^a,\label{npricci4}\\
\Phi_{02}&=\bar\Phi_{20}=\frac{1}{2}\Pi_{ab}m^am^b,\label{npricci5}\\
\Phi_{12}&=\bar\Phi_{21}=-\frac{1}{2\sqrt{2}}(\Pi_a+Q_a)\bar m^a.\label{npricci6}
\end{align}

Our expressions differ slightly from those reported in Ref.\cite{prat1}. In particular, we find a sign difference in $\Phi_{01}$, while the expressions for $\Phi_{00}$, $\Phi_{11}$ and $\Phi_{22}$ differ by additional terms and normalization factors.

\subsection{Weyl NP scalars}

The Weyl N--P scalars are calculated as
\begin{align}
\Psi_0&=(\mathcal{E}_{ab}-\tilde\eta_{ac}\mathcal{H}^c_{\ b})m^am^b=(\mathcal{E}_{ab}+\iu\mathcal{H}_{ab})m^am^b,\label{npweyl1}\\
\Psi_1&=-\frac{1}{\sqrt{2}}(\mathcal{E}_a-\tilde\eta_{ab}\mathcal{H}^b)m^a\nonumber\\
&=-\frac{1}{\sqrt{2}}(\mathcal{E}_a+\iu\mathcal{H}_a)m^a,\label{npweyl2}\\
\Psi_2&=\frac{1}{2}(\mathcal{E}-\iu\mathcal{H}),\label{npweyl3}\\
\Psi_3&=\frac{1}{\sqrt{2}}(\mathcal{E}_a+\iu\mathcal{H}_a)\bar m^a,\label{npweyl4}\\
\Psi_4&=(\mathcal{E}_{ab}+\tilde\eta_{ac}\mathcal{H}^c_{\ b})\bar m^a\bar m^b=(\mathcal{E}_{ab}+\iu\mathcal{H}_{ab})\bar m^a\bar m^b.\label{npweyl5}
\end{align}

The above scalars have also been computed in \cite{prat1}, although now there are sign modifications to \eqref{npweyl1} and \eqref{npweyl5}.

The Weyl scalars $\Psi_0,\cdots,\Psi_4$ are simply different projections of the Weyl tensor from those obtained by decomposing the electric and magnetic parts of the curvature into scalars, sheet vectors, and projected symmetric trace-free sheet tensors. The difference is therefore one of representation rather than geometric content.

In Table \ref{tab:NP122dictionary}, we highlight some key N--P scalars and their corresponding geometric interpretations in the 1+1+2 formulation.
\begin{table*}[ht!]
\centering
\renewcommand{\arraystretch}{1.1}
\begin{tabular}{c c c}
\hline
\textbf{NP quantity} & \textbf{\(1+1+2\) expression} & \textbf{Interpretation} \\
\hline

\(2\Psi_{2}\) 
& \(\mathcal{E}-i\mathcal{H}\) 
& Coulomb Weyl curvature \\

\(4\Phi_{00}\) 
& \(\rho+p+\Pi-2Q\) 
& Ingoing null matter flux \\

\(4\Phi_{22}\) 
& \(\rho+p+\Pi+2Q\) 
& Outgoing null matter flux \\

\(2\Re(\tilde{\rho})\) 
& \(-\theta_{(\ell)}\) 
& Outgoing null expansion \\

\(2\Re(\tilde{\mu})\) 
& \(\ \ \ \theta_{(n)}\) 
& Ingoing null expansion \\

\(\sqrt{2}\tilde{\sigma}\) 
& \(-
(\Sigma_{ab}+\zeta_{ab})m^{a}m^{b}\) 
& Shear of outgoing null congruence \\

\(\sqrt{2}\tilde{\lambda}\) 
& \(-
(\Sigma_{ab}-\zeta_{ab})\bar{m}^{a}\bar{m}^{b}\) 
& Shear of ingoing null congruence \\

\(\tilde{\tau}\) 
& \(-\tilde{\kappa}+(\Sigma_{a}+\eta_{ab}\Omega^{b})m^{a}\) 
& Twist/transport of null congruences \\

\(2\sqrt{2}\tilde{\epsilon}\) 
& \(
A+\left(\dfrac13\theta+\Sigma\right)
+(\dot{m}^{a}+\hat{m}^{a})\bar{m}_{a}
\) 
& Outgoing null inaffinity \\

\(2\sqrt{2}\tilde{\gamma}\) 
& \(
A-\left(\dfrac13\theta+\Sigma\right)
+(\dot{m}^{a}-\hat{m}^{a})\bar{m}_{a}
\) 
& Ingoing null inaffinity \\

\hline
\end{tabular}
\caption{Dictionary between selected NP quantities and the corresponding 1+1+2 covariant variables. The table highlights the geometric interpretation of several key NP scalars and spin coefficients in the 1+1+2 semi-tetrad formalism.}
\label{tab:NP122dictionary}
\end{table*}

\subsection{Some discussions on the gauge structure correspondence}

Here, we summaraze the relationship between the gauge structures of the two formalisms.

As have already been alluded to in the introduction, although they employ different bases and different sets of geometric variables, they describe the same spacetime geometry. The distinction between the two approaches lies primarily in the amount of frame gauge freedom retained. The NP formalism preserves the full local Lorentz gauge symmetry associated with a null tetrad, whereas the 1+1+2 formalism may be regarded as a partially gauge-fixed formulation in which a preferred timelike congruence and a preferred spatial direction are specified from the outset. Consequently, the 1+1+2 formalism retains only the residual gauge freedom corresponding to local rotations within the two-dimensional sheet.

In the NP formalism, one introduces a null tetrad
\begin{align}
\{\ell^a,n^a,m^a,\bar m^a\},
\end{align}
with normalizations as outlined at the beginning of the section. The absence of a preferred timelike or a preferred spatial direction implies that the full six-parameter Lorentz group acts freely on the tetrad. That is, retains the full local $SO(1,3)$ (or its universal cover) gauge symmetry, where the spin coefficients represent the associated connection one-forms in a null frame, i.e. the corresponding gauge transformations consist of spin-boost transformations together with null rotations about either $\ell^a$ or $n^a$. These transformations alter the individual spin coefficients and Weyl scalars while leaving the spacetime metric unchanged.

In contrast, the 1+1+2 formalism begins by choosing a unit timelike vector field $u^a$, representing a preferred family of observers. This immediately reduces the local Lorentz symmetry to the spatial rotation group $SO(3)$, since only transformations preserving $u^a$ are permitted. A further choice of a unit spacelike vector $e^a$ singles out a preferred spatial direction and reduces the residual symmetry to the subgroup $SO(2)$ acting on the sheet. That is, the fixing of the frame here yields the following reductions:
\begin{align}
SO(1,3)\longrightarrow SO(3)\longrightarrow SO(2),
\end{align}
which shows that the 1+1+2 formalism is obtained by progressively fixing the Lorentz frame.

{Comparing their gauge transformations}: The NP spin-boost transformation
\begin{align}
\ell^a\rightarrow f\ell^a,\quad n^a\rightarrow f^{-1}n^a,\quad m^a\rightarrow e^{i\chi}m^a,
\end{align}
where $f>0$ is smooth, and $\chi$ is smooth and denotes the phase angle of the NP tetrad parametrizing the $SO(2)$ rotation in the sheet, splits naturally into two distinct transformations in the 1+1+2 picture. Writing $f=e^{\varsigma}$, the boost acts as
\begin{align}
u^a&\rightarrow \cosh(\varsigma) u^a+\sinh(\varsigma) e^a,\\
e^a&\rightarrow \sinh(\varsigma) u^a+\cosh(\varsigma) e^a,
\end{align}
which is simply the Lorentz boost that mixes the timelike and radial directions. The phase rotation
\begin{align}
m^a\rightarrow e^{i\chi}m^a
\end{align}
corresponds to a local $SO(2)$ rotation of the sheet basis,
\begin{align}
  \left( {\begin{array}{c}
    e_3'\\
    e_4' \\
  \end{array} } \right)=\bar{\mathcal{R}}(\chi)
  \left( {\begin{array}{c}
    e_3 \\
    e_4\\
  \end{array} } \right),
\end{align}
where $\bar{\mathcal{R}}(\chi)\in SO(2)$. Thus the NP spin transformation is precisely the residual gauge symmetry of the 1+1+2 formalism.

The remaining NP gauge transformations, namely the null rotations about $\ell^a$ and $n^a$, do not possess independent analogues within a fully adapted semitetrad. A null rotation changes the null directions while preserving one of them, thereby simultaneously altering the corresponding combinations of $u^a$ and $e^a$. In the 1+1+2 formalism these transformations amount to changing the chosen timelike congruence or the preferred spatial direction. Since these vectors are fixed geometrically -- typically by symmetry, by the physical observer, or by the foliation -- such transformations are eliminated as gauge freedoms. Consequently, the absence of null rotations in the semitetrad formalism is not a deficiency but rather reflects the prior reduction of the Lorentz gauge.

These distinctions explain how the strengths of each approach complement the other. The NP formalism is advantageous when the physically preferred directions are null, as in gravitational radiation or asymptotic analyses at null infinity, where retaining the full Lorentz gauge provides considerable flexibility. By contrast, the 1+1+2 formalism is particularly effective whenever the spacetime possesses a naturally preferred timelike congruence and spatial direction, such as in spherically symmetric, locally rotationally symmetric, or horizon-adapted spacetimes. In these situations, the geometric symmetry itself fixes most of the Lorentz gauge, leaving only the intrinsic $SO(2)$ (rotation) freedom of the sheet. As a result, many variables vanish identically on the background and become automatically gauge invariant under perturbations, substantially simplifying especially physical interpretations. The complementarity means that we can adaptably switch between the formalisms depending on the nature of a problem in any particular spacetime or class of spacetimes.

From this perspective, the 1+1+2 semitetrad formalism is best understood as a geometrically adapted, partially gauge-fixed realization of the Newman–Penrose formalism. Both arise from the same Lorentz-gauge theory and encode the same connection and curvature, but they organize these quantities according to different geometric decompositions. Their gauge structures therefore differ not because they describe different physics, but because they retain different amounts of the underlying local Lorentz symmetry. This correspondence provides a unified geometric understanding of the two formalisms and clarifies why each is naturally suited to different classes of problems in mathematical relativity and gravitational physics.


\section{Brief comments on some implications for black hole horizons}\label{4}


In this section, we briefly illustrate how the relations established above may be used to characterize black hole horizons in terms of NP quantities. As a simple demonstration, we restrict attention to the class of locally rotationally symmetric (LRS) spacetimes, specifically the LRS class II subclass, which naturally admits the 1+1+2 decomposition \cite{chris1,chris2}. Owing to the underlying symmetry, all sheet vectors and projected trace-free tensors vanish identically, and the spacetime is completely described by covariantly defined scalar quantities. For LRS class II spacetimes, the vorticities \(\Omega\) and \(\xi\) also vanish.

One of the principal advantages of the \(1+1+2\) formalism is that it allows the Einstein field equations to be analyzed directly at the level of the preferred two-surfaces. A particularly important invariant associated with these surfaces is the Gaussian curvature scalar $\mathcal{K}$. In LRS class II spacetimes, $\mathcal{K}$ is related to the matter variables, the electric Weyl scalar, and the null expansions through
\begin{align}
\mathcal{K}=\frac{1}{3}(\varrho+\Lambda)-\left(\mathcal{E}+\frac{1}{2}\Pi\right)-\frac{1}{2}\theta_{(\ell)}\theta_{(n)},\label{gauss1}
\end{align}
where $\theta_{(\ell)}=-2\Re(\tilde\rho)$ and $\theta_{(n)}=2\Re(\tilde\mu)$, as defined in \eqref{expan1} and \eqref{expan2}, denote the expansions associated with the outgoing and ingoing null congruences, respectively. Rearranging \eqref{gauss1} gives
\begin{align}
\mathcal{E}-\frac{1}{3}\varrho+\frac{1}{2}\Pi=\frac{1}{3}\Lambda-\mathcal{K}-\frac{1}{2}\theta_{(\ell)}\theta_{(n)}.\label{gaussn1}
\end{align}
It will be assumed throughout that the local energy density $\varrho$ is nonnegative. Now, the directional derivatives of the left hand side of \eqref{gaussn1} along $u^a$ and $e^a$ are given respectively by \cite{chris2}
\begin{align}
\dot{\mathcal{E}}-\frac{1}{3}\dot{\varrho}+\frac{1}{2}\dot{\Pi}&=\left(\frac{2}{3}\theta-\Sigma\right)\left(-\frac{3}{2}\mathcal{E}-\frac{1}{4}\Pi+\frac{1}{2}(\varrho+p)\right)\nonumber\\
&+\frac{1}{2}\phi Q,\label{ev1}\\
\mathcal{E}'-\frac{1}{3}\varrho'+\frac{1}{2}\Pi'&=-\frac{1}{2}\left(\frac{2}{3}\theta-\Sigma\right)Q\nonumber\\
&-\frac{3}{2}\phi\left(\mathcal{E}+\frac{1}{2}\Pi\right).\label{pr1}
\end{align}
Taking the corresponding derivatives of \eqref{gaussn1} and then combining the resulting expressions yields
\begin{align}
&\sqrt{2}\theta_{(\ell)}\mathcal{Z}_1+\phi\mathcal{Z}_2\nonumber\\
&=-2\sqrt{2}\left(\mathcal{L}_{\ell}\mathcal{K}+\frac{1}{4}(\theta_{(n)}\mathcal{L}_{\ell}\theta_{(\ell)}+\theta_{(\ell)}\mathcal{L}_{\ell}\theta_{(n)}\right),\label{sol1}
\end{align}
where $\mathcal{L}_{\ell}$ denotes the Lie derivative along $\ell^a$, and 
\begin{align*}
\mathcal{Z}_1&=-Q-2\mathcal{E}+\mathcal{K}-\frac{2}{3}\varrho-p-\frac{1}{3}\Lambda+\frac{1}{4}\theta_{(\ell)}\theta_{(n)},\\
\mathcal{Z}_2&=2(Q+\mathcal{E}+\mathcal{K})+\left(\frac{5}{3}\varrho+p\right)+\frac{4}{3}\Lambda+\frac{1}{2}\theta_{(\ell)}\theta_{(n)}.
\end{align*}
For LRS class II spacetimes, the Gaussian curvature evolves according to
\begin{align}
\dot{\mathcal{K}}=-\left(\frac{2}{3}\theta-\Sigma\right)\mathcal{K},\quad \mathcal{K}'=-\phi\mathcal{K}.
\end{align}
from which it follows that
\begin{align}
\sqrt{2}\mathcal{L}_{\ell}\mathcal{K}=-\theta_{(\ell)}\mathcal{K}.
\end{align}

As we are interested in horizons, let us introduce the notion of a marginally outer trapped surface (MOTS) which generalizes cross sections of horizons. Consider a closed surface $\mathcal{S}$ in the decomposition under consideration. Then, the surface $\mathcal{S}$ is said to be marginally outer trapped if
\begin{align}
\theta_{(\ell)}=0
\end{align}
everywhere on \(S\). Assuming that such surfaces foliate a three-surface \(\mathcal{T}\), which is referred to as a marginally outer trapped tube (MOTT), one may define an everywhere tangent vector to $\mathcal{T}$
\begin{align}
x^a=\ell^a-Cn^a,
\end{align}
where the scalar function \(C\) determines the causal character of the horizon. In spherical symmetry,
\begin{align}
C=\frac{\mathcal{L}_{\ell}\theta_{(\ell)}}{\mathcal{L}_n\theta_{(\ell)}},\label{ccc}
\end{align}
since \(\mathcal{L}_{x}\theta_{(\ell)}=0\) along the horizon \cite{hay1,ib1}. In this case the sign of $C$ is constant over $\mathcal{T}$ and $\mathcal{T}$ is a horizon (see the reference \cite{ib2}). A spacelike horizon corresponds to \(C>0\) (a dynamical horizon), whereas \(C=0\) characterizes an isolated horizon. In context of LRS spacetimes the reader is referred to \cite{abb1,abb2} for explicit forms of the expression for $C$. 

We consider a future outer trapping horizon (FOTH) which contains trapped surfaces just to the inside, since we are interested in black hole horizons. In this case the ``outer'' is usually dropped and a MOTS is simply a MTS. For a future outer trapping horizon (FOTH), one requires
\begin{align}
\theta_{(n)}<0,
\quad
\mathcal{L}_{n}\theta_{(\ell)}<0.
\end{align}

Imposing the MOTS condition $\theta_{(\ell)}=0$, the constraint equation \eqref{sol1} reduces to
\begin{align}
2\phi\mathcal{Z}_2=-\sqrt{2}\theta_{(n)}\mathcal{L}_{\ell}\theta_{(\ell)}.\label{sol2}
\end{align}
We emphasize that all expressions are evaluated at the horizon. Using the relations derived in Section \ref{3}, this expression may be rewritten entirely in terms of NP quantities:
\begin{align}
(\Phi_{00}+\mathcal{Z}_2)\tilde\mu=0.\label{sol3}
\end{align}

Of course, for the case of a minimal cross section, i.e., $\tilde\mu=0$, the above is identically verified. Otherwise, for a FOTH we have
\begin{align}
\Phi_{00}+\mathcal{Z}_2=0.\label{sol4}
\end{align}
Assuming the null energy condition, $\Phi_{00}\geq0$, equation \eqref{sol4} implies that 
\begin{align}
\mathcal{Z}_2\leq0.
\end{align} 
Using \eqref{npricci1} and \eqref{npricci3}, we have $\Phi_{22}-\Phi_{00}=Q$, and we may write $\mathcal{Z}_2$ in terms of the NP scalars
\begin{align}
\mathcal{Z}_2&=2\mathcal{F}+\left(\frac{5}{3}\varrho+p\right),\label{z2np}
\end{align}
where
\begin{align*}
\mathcal{F}=(\Phi_{22}+2\Psi_2+\mathcal{K})+\frac{2}{3}\Lambda.
\end{align*}
If one further imposes the strong energy condition (SEC), which implies the NEC $\varrho+3p\geq0$, then the parenthesized term of \eqref{z2np} is positive, and since $\mathcal{Z}_2\leq0$, necessarily 
\begin{align}
\mathcal{F}\leq0.
\end{align} 
Substituting \eqref{z2np} into \eqref{sol4} yields
\begin{align}
\frac{1}{2}\Phi_{00}&=\mathcal{F}+\frac{1}{2}\left(\frac{5}{3}\varrho+p\right).\label{constraint1}
\end{align}
Therefore, we have that 
\begin{align}
\frac{1}{2}\Phi_{00}\geq\mathcal{F}.\label{sol5} 
\end{align}

Since we are considering black hole horizons (FOTH), spherical topology is necessary for the MOTS (this follows from a notion of stability for MOTS for which the interested reader is referred to the references \cite{and1,and2} for details). Therefore, the condition
\begin{align}
\Phi_{22}+2\Psi_2+\frac{2}{3}\Lambda\leq0,\label{sol6}
\end{align}
is both necessary and sufficient to guarantee the non-positivity of $\mathcal{F}$ under the SEC assumption. If \eqref{sol6} fails, $\mathcal{F}>0$ and this fails the SEC as $\mathcal{Z}_2$ is required to be non-positive. Equation \eqref{sol6} therefore provides a purely geometric criterion, expressed in terms of NP scalars, that constrains the existence of black hole horizons in LRS class II spacetimes. In particular, if this inequality fails in some region of spacetime satisfying the SEC, then that region cannot admit a future outer trapping horizon. Since \(\Psi_2\) encodes the Coulomb part of the free gravitational field together with contributions from the heat flux, the condition illustrates that horizon formation depends on a delicate balance between matter flux and Weyl curvature.

Consider, for example, a LRS II spacetime that is anisotropic but with vanishing heat flux ($Q=0,\Pi\neq0$). The scalars $\Phi_{00}$ and $\Phi_{22}$ coincide and SEC then imposes
\begin{align}
\Psi_2+\frac{1}{3}\Lambda\leq0.
\end{align}
Thus, in the absence of a cosmological constant, these spacetimes will not admit a FOTH for a positive $\Psi_2$. 

The case of a generic isolated horizon embedded in a vacuum configuration admits an especially simple characterization entirely reliant on the sign of the cosmological constant. For such a case, an important invariant characterizing the horizon geometry is the complex scalar \cite{ash1,lew1}
\begin{align}
\bar\psi=-\frac{1}{2}\mathcal{K}-\iu\Im(\Psi_2),\label{psi2_1}
\end{align}
where $\Im(\ )$ denotes ``the imaginary part of'', which in our formulation takes the form
\begin{align}
\bar\psi=-\frac{1}{2}(\mathcal{K}-\iu\mathcal{H}).\label{psi2_2}
\end{align}
This complex scalar relates to $\Psi_2$ via \cite{ash1,lew1}
\begin{align}
\Psi_2=\bar\psi+\frac{\Lambda}{6}=-\frac{1}{2}(\mathcal{K}-\iu\mathcal{H})+\frac{\Lambda}{6}.\label{psi2_3}
\end{align}
Comparing this with the relation obtained for $\Psi_2$ yields
\begin{align}
\mathcal{H}=0,\quad \frac{1}{3}\Lambda=\mathcal{E}+\mathcal{K}=2\Psi_2+\mathcal{K}.\label{iso1}
\end{align}
For LRS II solutions the vanishing condition on the magnetic Weyl scalar $\mathcal{H}$ is trivially met, and from the second condition of \eqref{iso1} and noting that $\Phi_{22}=0$ for vacuum, we have
\begin{align}
\mathcal{F}=\Lambda.
\end{align}
Therefore, the condition \eqref{sol6} now simply becomes
\begin{align}
\Lambda\leq0.\label{forih}
\end{align}
Thus, in a $\Lambda$-vacuum LRS II geometry satisfying the SEC, a positive $\Lambda$ obstructs the existence of an isolated horizon, except when the MOTS are minimal (in this case the equation \eqref{sol3} holds trivially). Of course, necessarily, $\Psi_2<0$ on the horizon by \eqref{iso1}.

In fact, it turns out that this horizon existence constraint from the cosmological constant may be generic for $\tilde\mu\neq0$, for LRS spacetimes: using the Gauss curvature, after some basic algebra, we may rewrite the function $\mathcal{F}$ in terms of a Ricci NP scalar, cosmological constant and curvature variables as
\begin{align}
2\mathcal{F}=2\Lambda-\Phi_{00}+\left(\frac{5}{3}\varrho+p\right).\label{constraint2}
\end{align}
which we compare to \eqref{constraint1} to obtain
\begin{align}
\Lambda=2\mathcal{F}.\label{constraint3}
\end{align}
Therefore, a positive $\Lambda$ implies $\mathcal{F}>0$. Thus, {\em for a LRS II geometry obeying the SEC, a positive $\Lambda$ obstructs the existence of a FOTH.} This result is true whether or not we are in a vacuum. Indeed, generally, if $\Lambda\leq0$, one has to go back and verify \eqref{sol6}.

Notice that on an isolated horizon, if the SEC strictly holds, $\mathcal{F}>\Lambda$. For these solutions then, a strictly negative $\Lambda$ is necessary for the existence of an FOTH. It follows that, for a LRS II geometry with a vanishing cosmological constant and obeying the SEC strictly, any FOTH in such geometry is spacelike.

These results demonstrate how the mapping between the NP and 1+1+2 formalisms provides a geometrically transparent framework for analyzing black hole horizons. More broadly, they suggest that several well-known horizon conditions expressed in the NP language may admit natural geometric reinterpretations within the covariant 1+1+2 approach.


\section{Summary and outlook}\label{5}


In this work, we have, to our knowledge, for the first time, derived the complete set of NP scalars in terms of the scalar, vector, and tensor quantities of the 1+1+2 semi-tetrad covariant formulation of general relativity. The resulting dictionary provides a translation tool between two complementary geometric descriptions of spacetime and assigns a direct covariant interpretation to all Newman--Penrose curvature scalars and spin coefficients.

The distinction between these approaches reflects the different geometric adaptations of the two formalisms. The NP formalism is best suited to situations where null congruences are the fundamental objects, whereas the 1+1+2 formalism is most effective when the spacetime possesses a preferred timelike congruence and a preferred spatial direction. Consequently, problems involving spherically symmetric backgrounds, locally rotationally symmetric cosmologies, gravitational collapse, and dynamical horizons often admit a much more economical treatment in the 1+1+2 framework. In these cases, the symmetry of the spacetime itself performs most of the gauge fixing, leaving a minimal residual $SO(2)$ freedom and allowing the field equations to be written directly in terms of geometrically meaningful, often automatically gauge-invariant, variables. This close alignment between the formalism and the underlying geometry is the principal reason why the 1+1+2 approach can be substantially more convenient than the Newman -- Penrose formalism in such applications.

Given the complementary strengths of these formalisms, we expect that this correspondence may prove useful in a variety of contexts within general relativity and relativistic astrophysics.

As a simple application of this mapping, we have considered conditions governing the existence of black hole horizons in the class of LRS II spacetimes. In particular, we obtained necessary conditions for the existence of dynamical and isolated horizons expressed purely in terms of the Ricci and Weyl NP scalars together with the cosmological constant, specifically \(\Phi_{22}\) and \(\Psi_2\). It is also found that the existence of a future outer trapping horizon in these spacetimes favors a non-positive cosmological constant. These conditions provide a simple criterion for ruling out the existence of black hole horizons in regions where the inequalities fail to hold.

There are several promising directions in which the mapping presented in this letter may be further exploited. For example, the horizon analysis outlined here could potentially be extended to more general classes of black hole horizons beyond the LRS setting. Another natural application concerns gravitational perturbation theory, particularly in light of the work of Pratten \cite{prat1} in the context of \(f(R)\) gravity. Recasting known perturbative results formulated in the NP approach in terms of the geometric variables of the 1+1+2 formalism may provide deeper insight into the interplay between spacetime geometry and gravitational perturbations.

Another interesting direction arises from the work of \cite{abb3}, where the dynamics of a null horizon subjected to linear perturbations was investigated. It is conceivable that the equations governing the perturbation dynamics of the horizon may admit a more transparent geometric interpretation, or perhaps even simplification, when reformulated within the present framework. Finally, it would be worthwhile to investigate the classification of general 1+1+2 spacetimes according to Petrov type, which would naturally encompass perturbations of locally rotationally symmetric geometries and could provide further insight into the geometric structure of perturbative spacetimes.


\begin{acknowledgments}


We thank the anonymous referee for the helpful suggestions to include discussions on the gauge correspondence and to add depth to the advantages of the correspondence. AS acknowledges that this research is supported by the Institute of Mathematics, funded through the High-level Talent Research Start-up Project Funding of the Henan Academy of Sciences (Project No.: 251819085). PKSD acknowledges support by grant from the First Rand Bank, South Africa.
\end{acknowledgments}


\end{document}